\documentclass[journal=jacsat,manuscript=article]{achemso}
\usepackage{chemformula}
\usepackage{graphicx}
\usepackage{dcolumn}
\usepackage{bm}
\usepackage{float}
\usepackage{mathtools}   
\usepackage{algorithm}
\usepackage{algpseudocode}
\usepackage{mathptmx}
\usepackage{amsmath,amsthm,amssymb}
\usepackage{siunitx}

\newcommand{\epsd}{\epsilon^\mathrm{ML}_\mathrm{d}} 
\newcommand{\epso}{\epsilon^\mathrm{ML}_\mathrm{o}}
\newcommand{\qmsbt}{QM7b-T}
\newcommand{\gdbt}{GDB-13-T}

\newcommand{\gprrfc}{RC/GPR/RFC}
\newcommand{\lrrfc}{RC/LR/RFC}

\newcommand\footnoteref[1]{\protected@xdef\@thefnmark{\ref{#1}}\@footnotemark}

\algnewcommand\Input{\item[\textbf{Input:}]}
\algnewcommand\Output{\item[\textbf{Output:}]}
\DeclareMathOperator*{\argmin}{arg\,min}

\newcommand{\R}{\mathbb{R}}

\author{Lixue Cheng}
\affiliation{
Division of Chemistry and Chemical Engineering, California Institute of Technology, Pasadena, CA 91125, USA
}
\author{Nikola B. Kovachki}
\affiliation{Computing and Mathematical Sciences, California Institute of Technology, Pasadena, CA 91125, USA}
\author{Matthew Welborn}
\affiliation{
Division of Chemistry and Chemical Engineering, California Institute of Technology, Pasadena, CA 91125, USA
}
\author{Thomas F. Miller III}
\email{tfm@caltech.edu}
\affiliation{
Division of Chemistry and Chemical Engineering, California Institute of Technology, Pasadena, CA 91125, USA
}

\title{Regression-clustering for Improved Accuracy and Training Cost with Molecular-Orbital-Based Machine Learning}
\date{\today}
             
\SectionNumbersOn 
\begin{document}

\begin{abstract}
Machine learning (ML) in the representation of molecular-orbital-based (MOB) features has been shown to be an accurate and transferable approach to the prediction of post-Hartree-Fock correlation energies. 
Previous applications of MOB-ML employed Gaussian Process Regression (GPR), which provides good prediction accuracy with small training sets; however, the cost of GPR training scales cubically with the amount of data and becomes a computational bottleneck for large training sets.
In the current work, we address this problem by introducing
a clustering/regression/classification implementation of MOB-ML.
In a first step, regression clustering (RC) is used to partition the training data to best fit an ensemble of linear regression (LR) models; in a second step, each cluster is regressed independently, using either LR or GPR; and in a third step, a random forest classifier (RFC) is trained for the prediction of cluster assignments based on MOB feature values.
Upon inspection, RC is found to recapitulate chemically intuitive groupings of the frontier molecular orbitals, and the combined RC/LR/RFC and RC/GPR/RFC implementations of MOB-ML are found to provide good prediction accuracy with greatly reduced wall-clock training times. 
For a dataset of thermalized (350 K) geometries of 7211 organic molecules of up to seven heavy atoms (\qmsbt{}),  both RC/LR/RFC and RC/GPR/RFC reach chemical accuracy (1 kcal/mol prediction error) with only 300 training molecules, while providing  
35000-fold and 4500-fold reductions in the wall-clock training time, respectively, compared to MOB-ML without clustering.
The resulting models are also demonstrated to retain transferability for the prediction of large-molecule energies with only small-molecule training data.
Finally, it is shown that capping the number of training datapoints per cluster leads to further improvements in prediction accuracy with negligible increases in wall-clock training time.
\end{abstract}

\section{Introduction}

Machine-learning (ML) continues to emerge as a versatile strategy in the chemical sciences, with applications to drug discovery,\cite{Lavecchia2015, Gawehn2016, Popova2018, kearnes2016molecular, Mater2019} materials design,\cite{Kim2017, Ren2018, Butler2018, Sanchez-Lengeling2018, Mater2019} and reaction prediction.\cite{Wei2016, Raccuglia2016, Ulissi2017,Segler2017, Segler2018, Mater2019} 
An increasing number of ML methods have focused on the prediction of molecular properties, including quantum mechanical electronic energies \cite{Smith2017, smiti_transfer_2018, Lubbers, Bartok2010, rupp2012fast, VonLilienfeld2013,hansen2013assessment, ramakrishnan2015big,Behler2016,Paesani2016,schutt2017quantum, wu2018moleculenet,Nguyen2018, Fujikake2018, Li2018,Zhang2018, Nandy2018}, densities \cite{wu2018moleculenet, Grisafi2019,Grisafi2019b,Bogojeski2018,  Pereira2018, Smith2019}, and spectra \cite{Ramakrishnan2015, Gastegger2017, Yao2018, Christensen2019, Ghosh2019}.  
 Most of this work has focused on ML in the representation of atom- or geometry-specific features, although more abstract representations are gaining increased attention.\cite{Tuckerman, mcgibbon2017improving,Takuro2019,Townsend2019,pande2019arxiv, Welborn2018,Cheng2019} 

We recently introduced a rigorous factorization of the post-Hartree-Fock correlation energy into contributions from pairs of occupied molecular orbitals and showed that these pair contributions could be compactly represented in the space of molecular-orbital-based (MOB) features to allow for straightforward ML regression.\cite{Welborn2018,Cheng2019}  
This MOB-ML method  was demonstrated to accurately predict second-order M{\o}ller-Plessett perturbation theory (MP2)\cite{Moller1934,LMP2} and coupled cluster with singles, doubles and perturbative triples (CCSD(T))\cite{Bartlett1990,LCCSDT} energies of different benchmark systems, including the \qmsbt{} and \gdbt{} datasets of thermalized drug-like organic molecules.  
While providing good accuracy with a  modest amount of training data, the accuracy of MOB-ML in these initial studies was limited by the high computational cost ($\mathcal{O}(N^3)$) of applying  Gaussian Process Regression (GPR) to the full set of training data.\cite{Cheng2019} 

In this work, we combine MOB-ML with  regression clustering (RC) to overcome this bottleneck in computational cost and accuracy. 
The training data are clustered via RC to discover locally linear structures.
By independently regressing these subsets of the data, we obtain MOB-ML models with greatly reduced training costs while preserving prediction accuracy and transferability.

\section{Theory}
\label{sec:theory}

\subsection{Molecular-orbital based machine learning (MOB-ML)}
The MOB-ML method is based on the observation that the correlation energy for any post-Hartree-Fock wavefunction theory can be exactly decomposed as a sum over occupied molecular orbitals (MOs) via Nesbet's theorem,\cite{Nesbet1958,SzaboNesbet} 
\begin{equation}
\label{Ecorr}
E_\textrm{c} 
=\sum^{\textrm{occ}}_{ij}\epsilon_{ij},
\end{equation}
where $E_c$ is the correlation energy and $\epsilon_{ij}$ is the pair correlation energy corresponding to occupied MOs $i$ and $j$. 
The pair correlation energies can be expressed as a functional of the set of (occupied and unoccupied) MOs, appropriately indexed by $i$ and $j$, such that
\begin{equation}
    \epsilon_{ij} = \epsilon\left[\{\phi_p\}^{ij} \right].
\end{equation}
The functional $\epsilon$ maps the Hartree-Fock MOs to the pair correlation
energy, regardless of the molecular composition or geometry, such that it is a universal functional for all chemical systems.
To bypass the expensive post-Hartree-Fock evaluation procedure, MOB-ML approximates $\epsilon_{ij}$ by machine learning two functionals, $\epsd$ and $\epso$, which correspond to diagonal and off-diagonal terms of the sum in Eq. \ref{Ecorr}. 

\begin{equation}
    \label{eq:diag_and_offdiag}
    \epsilon_{ij} \approx
    \begin{cases}
        \epsd \left[\mathbf{f}_i\right] & \text{if $i=j$} \\
        \epso \left[\mathbf{f}_{ij}\right] & \text{if $i\ne j$}
    \end{cases}
\end{equation}
The MOB-ML feature vectors $\mathbf{f}_i$ and $\mathbf{f}_{ij}$ are comprised of unique elements of the Fock, Coulomb and exchange matrices between $\phi_i$, $\phi_j$, and the set of virtual orbitals.
Without loss of generality, we perform MOB-ML using localized MOs (LMOs) to improve transferability across chemical systems.\cite{Welborn2018}
Detailed descriptions of feature design are provided in our previous work \cite{Cheng2019, Welborn2018}, and the features employed here are unchanged from those detailed in Ref.~\citenum{Cheng2019}. 

\subsection{Local linearity of MOB feature space}
\label{sec:theory:motivate}

It has been previously emphasized that MOB-ML facilitates transferability across chemical systems, even allowing for predictions involving molecules with  elements that do not appear in the training set,\cite{Welborn2018}  due to the fact that MOB features provide a compact and highly abstracted representation of the electronic structure.  However, it is worth additionally emphasizing that this transferability benefits from the smooth variation and local linearity of the 
pair correlation energies as a function of  MOB feature values associated with different molecular geometries and even different molecules.  

\begin{figure}[htbp]
\includegraphics[width=0.5\columnwidth]{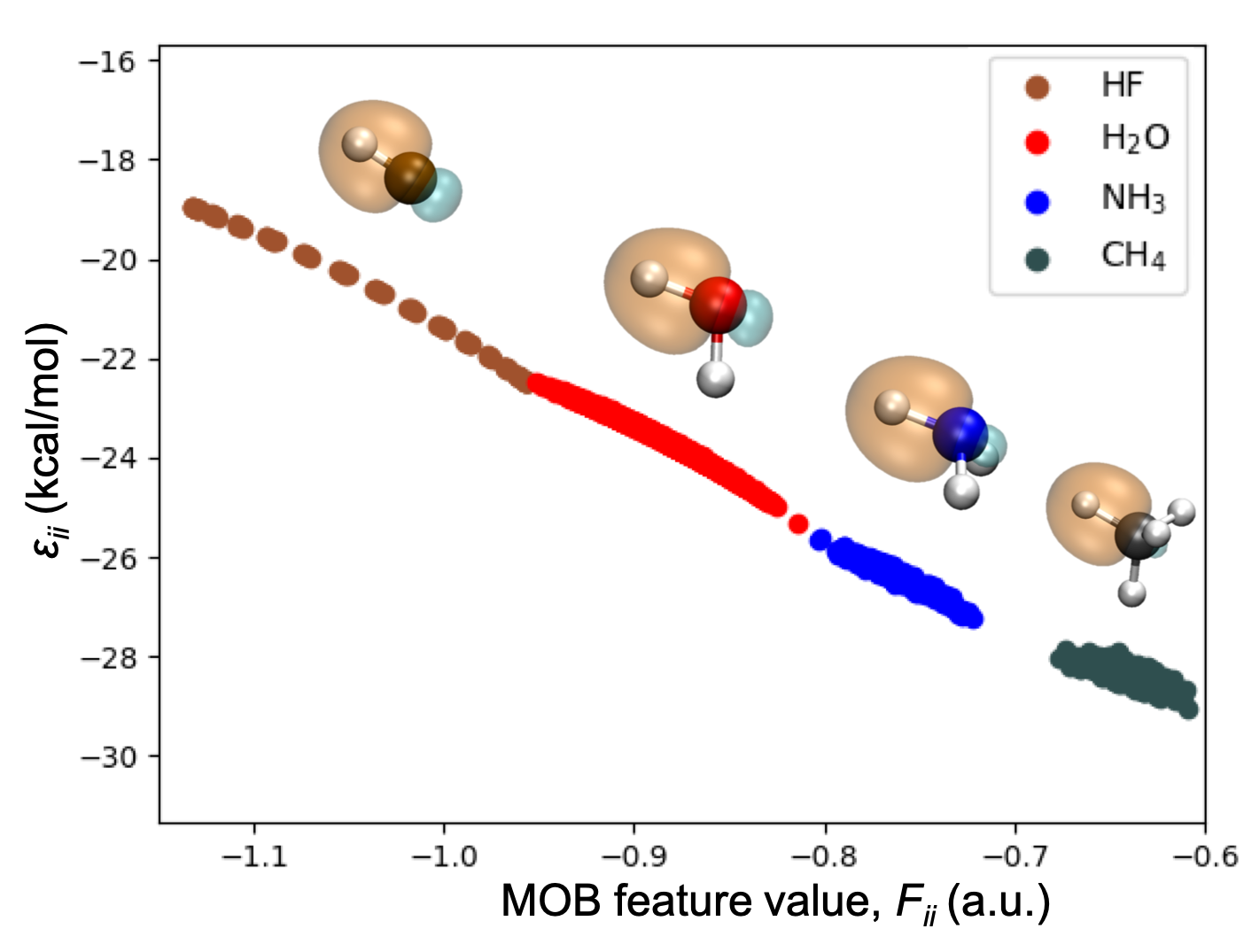}
\caption{The diagonal pair correlation energy ($\epsilon_{ii}$) 
for a localized $\sigma$-bond in four different molecules at thermally sampled geometries (at 350 K), computed at the MP2/cc-pvTZ level of theory.
The diagonal pair correlation energies for \ch{HF}, \ch{NH3}, and \ch{CH4} are shifted vertically downward relative to those of \ch{HF}  by 3.407, 6.289, and 7.772 kcal/mol for \ch{H2O}, \ch{NH3} and \ch{CH4}. Illustrative $\sigma$-bond LMOs are shown for each molecule.}
\label{figure:sigma_linear}
\end{figure}

Figure \ref{figure:sigma_linear} illustrates these latter properties for a $\sigma$-bonding orbital in a series of simple molecules. On the y-axis, we plot the diagonal contribution to the correlation energy associated with this orbital ($\epsilon_{ii}$), computed at the MP2/cc-pvTZ level of theory. On the x-axis, we plot the value of a particular MOB feature, the Fock matrix element for the that localized orbital, $F_{ii}$. For each molecule, a range of geometries is sampled from the Boltzmann distribution at $350$ K, with each plotted point corresponding to a different sampled geometry.

It is immediately clear from the figure that the pair correlation energy varies smoothly and linearly as a function of the MOB feature value. Moreover, the slope of the linear curve is remarkably consistent across molecules.  This  illustration suggests that MOB features may lead to accurate regression of correlation energies using  simple machine learning models (even linear models), and it also indicates the basis for the robust transferability of MOB-ML across diverse chemical systems, including those with elements that do not appear in the training set.

\subsection{Regression clustering  with a greedy algorithm}
\label{sec:theory:method}

To take advantage of the local linearity of pair correlation energies as a function of MOB features, we propose a strategy to discover  optimally linear clusters using regression clustering (RC).\cite{spath79} Consider the set of $M$ datapoints  \{$\mathbf{f}_{t},\epsilon_{t}$\} $\subset \R^d \times \R$, where $d$ is the length of the MOB feature vector and where each datapoint is indexed by $t$ and corresponds to a MOB feature vector and the associated reference value (i.e., label) for the pair correlation energy. To separate these datapoints into locally linear clusters, \(S_1,\dots,S_N\), we seek a solution to the optimization problem
\begin{equation}
\label{eq:optimization_problem}
\min_{S_1,\dots,S_N} \sum_{k=1}^N \sum_{t \in S_k} |A(S_k) \cdot \mathbf{f}_{t} + b(S_k) - \epsilon_{t}|^2
\end{equation}
where \(A(S_k) \in \R^d\) and \(b(S_k) \in \R\) are obtained via ordinary least squares (OLS) solution, 
\begin{equation}
\label{eq:ols_problem}
\begin{bmatrix} 
\mathbf{f}_{t_1}^T & 1 \\
\vdots & \vdots \\
\mathbf{f}_{t_{|S_k|}}^T & 1
\end{bmatrix}
\begin{bmatrix}
A(S_k) \\
b(S_k)
\end{bmatrix}
= 
\begin{bmatrix}
\epsilon_{t_1} \\
\vdots \\
\epsilon_{t_{|S_k|}}
\end{bmatrix}.
\end{equation}
Each resulting $S_k$ is the set of indices $t$ assigned to cluster $k$ comprised of $|S_k|$ datapoints. 
To perform the optimization in Eq.~\ref{eq:optimization_problem}, we employ a modified version of the greedy algorithm proposed in Ref.~\citenum{RC} (Algorithm~\ref{alg:greedy}). In general, solutions to Eq. \ref{eq:optimization_problem} may overlap, such that \(S_k \cap S_l \neq \emptyset\)
for \(k \neq l\); however, the proposed algorithm enforces that clusters remain pairwise-disjoint. 

\begin{algorithm}[h]
\caption{Greedy algorithm for the solution of Eq. \ref{eq:optimization_problem}.}
\label{alg:greedy}
    \begin{algorithmic}[1]
        \Input{Initial clusters: $S_1,\dots,S_N$}
        \Output{Data clusters $S_1,\dots,S_N$}
        \For{$k \gets 1$ to $N$}
        \State $A(S_k), b(S_k) \gets$ OLS solution of
        Eq. \ref{eq:ols_problem}
        \EndFor
        \While{not converged}
            \For{$k \gets 1$ to $N$}
            \State 
            $S_k \gets \{ t \in \{1,\dots,M\}: \argmin\limits_{n \in \{1,\dots,N\}} |A(S_n) \cdot \mathbf{f}_{t} + b(S_n) - \epsilon_{t}|^2 = k\}$
            \label{line6}
            \EndFor
            \For{$ k \gets 1$ to $N$}
            \State $A(S_k), b(S_k) \gets $ OLS solution of Eq. \ref{eq:ols_problem}
            \EndFor
        \EndWhile
    \end{algorithmic}
\end{algorithm}

Algorithm \ref{alg:greedy} has a per-iteration runtime of \(\mathcal{O}(Md^2)\),
since we compute \(N\) OLS solutions each with runtime \(\mathcal{O}(|S_k|d^2)\)
and since \(\sum_{k=1}^N |S_k| = M\).  However, the algorithm can be trivially parallelized to reach a runtime of \(\mathcal{O}(\max(|S_k|)d^2)\). A key operational step in this algorithm is line \ref{line6}, which can be explained in simple terms as follows:  we assign each datapoint, indexed by $t$, to the cluster to which it is closest, as measured by the squared linear regression distance metric, 
\begin{equation}
|D_{n,t}|^2=|A(S_n) \cdot \mathbf{f}_{t} + b(S_n) - \epsilon_{t}|^2
\label{distance_metric}
\end{equation} 
where $D_{n,t}$ is the distance of this point to cluster $n$.
In principle, a datapoint could be equidistant to two or more different clusters by this metric; in such cases, we  randomly assign the datapoint to only one of those equidistant clusters to enforce the  pairwise-disjointness of the resulting clusters. Convergence of the greedy algorithm is measured by the decrease in the objective function of Eq.~\ref{eq:optimization_problem}.

\begin{figure}[htbp]
\includegraphics[width=0.9\columnwidth]{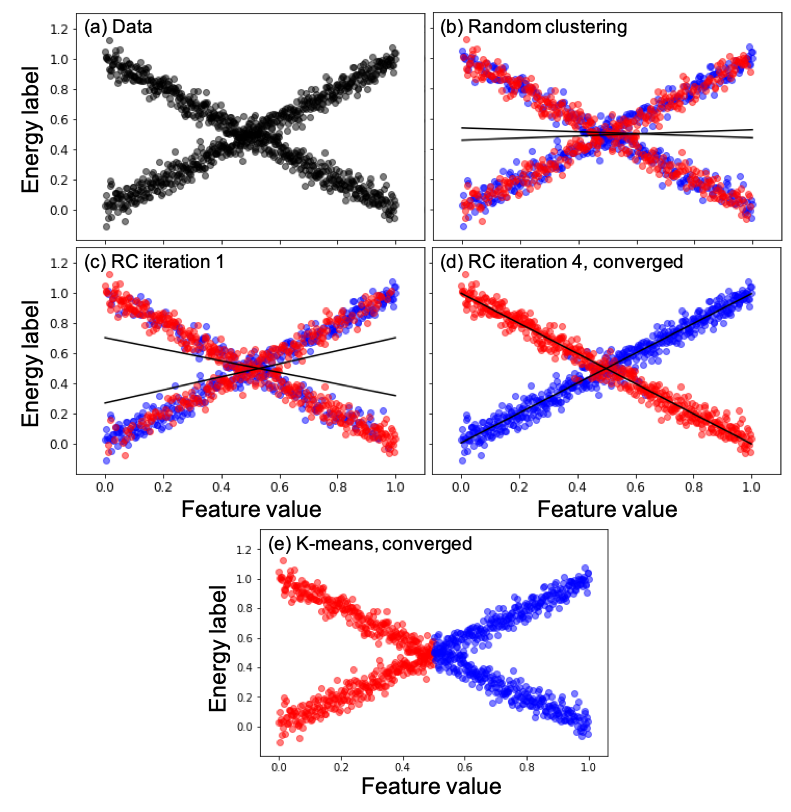}
\caption{Comparison of clustering algorithms for (a) a dataset comprised of two cluster of nearly linear data that overlap in feature space, using  (b-d) RC and (e) standard K-means clustering. (b) Random initialization of the clusters for the greedy algorithm, with datapoint color indicating  cluster assignment.  (c) Cluster assignments after one iteration of the greedy algorithm.  (d) Converged cluster assignments after four iterations of the greedy algorithm. For panels (b-d), two  linear regression lines at each iteration are shown in black. 
(e) Converged cluster assignments obtained using K-means clustering, which fails to reveal the underlying linear structure of the clusters.}
\label{fig:greedy}
\end{figure}

Figure \ref{fig:greedy} illustrates RC in a simple one-dimensional example for which   unsupervised clustering 
approaches will fail to reveal the underlying linear structure. 
To create two  clusters of nearly linear data that overlap in feature space, the interval of feature values on \([0,1]\) is uniformly discretized, such that \(\mathbf{f}_{t} = (t-1)/(M-1)\) for \(t=1,\dots,M\). Then, $M/2$ of the feature values are randomly chosen without replacement for cluster \(S_1\) while the remainder are placed in \(S_2\); the energy labels associated with each feature value are then generated using \[\epsilon_{t} = \mathbf{f}_{t} + \xi_{t,1}, \quad t \in S_1\]
and 
\[\epsilon_{t} = -\mathbf{f}_{t} + 1 + \xi_{t,2}, \quad t \in S_2\]
where \(\xi_{t,k} \sim \mathcal{N}(0,0.1^2)\) is an i.d.d.~sequence. The resulting dataset is shown in Fig.~\ref{fig:greedy}a. 

Application of the RC method to this example is illustrated in Figs.~\ref{fig:greedy}(b-d).
The greedy algorithm is initialized by randomly assigning each datapoint to either \(S_1\) or \(S_2\) (Fig.~\ref{fig:greedy}b).
Then, with only a small number of iterations (Figs.~\ref{fig:greedy}c and d), the algorithm converges to clusters that reflect the underlying linear character. 
For comparison, Fig.~\ref{fig:greedy}e shows the clustering that is obtained upon convergence of the standard K-means algorithm,\cite{kmeans} initialized with random cluster assignments.  Unlike RC, the K-means algorithm prioritizes the compactness of clusters, resulting in a final clustering that is far less amenable to simple regression.   
While we recognize that the correct clustering could potentially be obtained using K-means when the dimensions of \(\mathbf{f}_{t}\) and \(\epsilon_{t}\) are comparable, this is  not the case for MOB-ML applications since \(\mathbf{f}_{t}\) is typically at least 10-dimensional and \(\epsilon_{t}\) is a scalar; the RC approach does not suffer from this issue.
Finally, we have confirmed that initialization of RC from the clustering in Fig.~\ref{fig:greedy}e rapidly returns to the results in Fig.~\ref{fig:greedy}d, requiring only a couple of iterations of the greedy algorithm.

\begin{figure*}[htbp]
\includegraphics[width=0.95\textwidth]{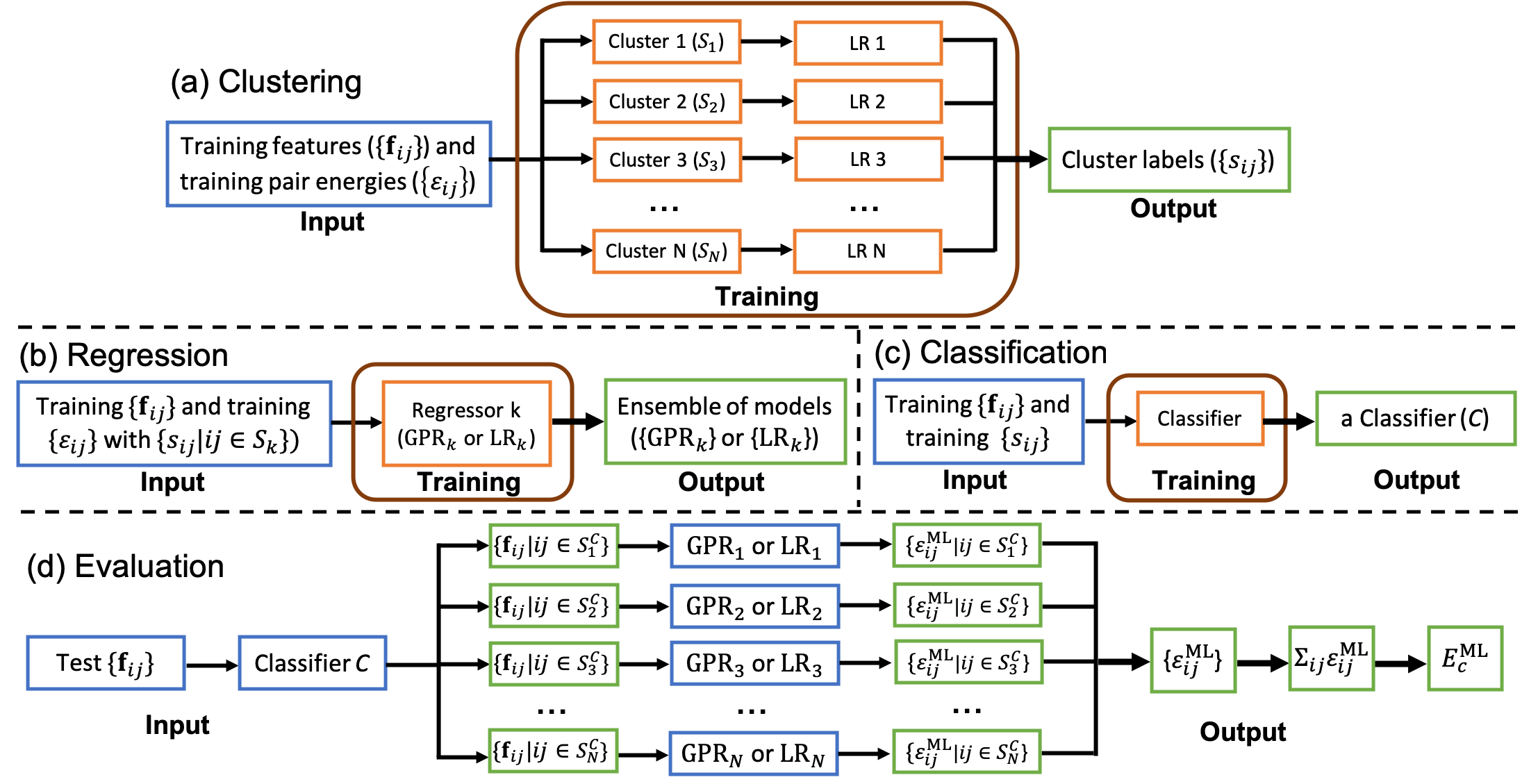}
\caption{The MOB-ML clustering/regression/classification workflow. (a) Clustering of the
training dataset of MOB-ML feature vectors and energy labels using RC to  obtain  optimized  linear clusters and to provide the cluster labels for the feature vectors. 
(b) Regression of each cluster of training data (using LR or GPR), to obtain the ensemble of cluster-specific regression models. 
(c) Training a classifier (RFC) from the MOB-ML feature vectors and cluster labels for the training data. 
(d) Evaluating the predicted MOB-ML pair correlation energy from a test feature vector is performed by first classifying the feature vector into one of the clusters, then evaluating the cluster-specific regression model.
In each panel, blue boxes indicate input quantities,
orange boxes indicate training intermediates, and green boxes indicate the resulting labels, models and pair correlation energy predictions. }
\label{fig:diagram}
\end{figure*}

\section{Calculation Details}
\label{sec:details}

Results are presented for \qmsbt{},\cite{Cheng2019} a thermalized version of the QM7b set\cite{QM7b} of 7211 molecules with up to seven C, O, N, S, and Cl heavy atoms, as well as for \gdbt,\cite{Cheng2019} a thermalized version of the GDB-13 set\cite{GDB-13} of molecules with thirteen C,
O, N, S, and Cl heavy atoms. The MOB-ML features employed in the current study are identical to those previously provided.\cite{Cheng2019}
Reference pair correlation energies are computed using MP2\cite{Moller1934} and using
CCSD(T).\cite{Bartlett1990,LCCSDT}
The MP2 reference data were obtained with the cc-pVTZ basis set,\cite{Dunning1989} whereas the CCSD(T) data were obtained using the cc-pVDZ basis set.\cite{Dunning1989} 
All employed training and test datasets are provided in Ref.~\citenum{Cheng2019}.

\subsection{Regression Clustering (RC)} 
RC is performed using the ordinary least square linear regression implementation in the \textsc{Scikit-learn}  package \cite{scikit-learn}. Unless otherwise specified, we initialize the greedy algorithm from the results of K-means clustering, also implemented in \textsc{Scikit-learn}; K-means initialization was found to improve the subsequent training of the random forest classifier (RFC) in comparison to random initialization. It is found that neither L1 nor L2 regularization had significant effect on the rate of convergence of the greedy algorithm, so neither is employed in the results presented here. It is found that a convergence threshold of \num{1e-8} kcal$^2$/mol$^2$  for the loss function of the greedy algorithm  (Eq.~\ref{eq:optimization_problem}) leads to no degradation in the final MOB-ML regression accuracy (Fig.~S2); this value is employed throughout.

\subsection{Regression}
Two different regression models are employed in the current work.  The first is ordinary least-squares linear regression (LR), as implemented in \textsc{Scikit-learn}.  The second is Gaussian Process Regression, as implemented in the  \textsc{GPy} 1.9.6 software package \cite{gpy2014}. 
Regression is independently performed  for the training data associated with each cluster, yielding a local regression model for each cluster. Also, as in our previous work,\cite{Welborn2018, Cheng2019} regression is independently performed for the diagonal and off-diagonal pair correlation energies ($\epsd$ and $\epso$) yielding independent regression models for each (Eq. \ref{eq:diag_and_offdiag}). 

GPR is performed using a negative log marginal likelihood objective.  As in our previous work,\cite{Cheng2019} the Mat\'ern 5/2 kernel is used for regression of the diagonal pair correlation energies and the Mat\'ern 3/2 kernel is used for the off-diagonal pair correlation energies; in both cases, white noise regularization\cite{rasmussen2006} is employed, and the GPR is initialized with unit lengthscale and variance.  

\subsection{Classification}
An RFC is trained on MOB-ML features and cluster labels for a training set and then used to predict the cluster assignment of test datapoints in MOB-ML feature space.
We employ the RFC implementation in \textsc{Scikit-learn}, using with 200 trees, the entropy split criteria,\cite{criminisi2012decision}
and balanced class weights.\cite{criminisi2012decision} 
Alternative classifiers were also tested in this work, including K-means, Linear SVM,\cite{lsvm} and AdaBoost;\cite{adaboost} however, these schemes were generally found to yield less accurate MOB-ML energy predictions than RFC.

For comparison, a ``perfect" classifier is obtained by simply including the test data within the RC training set.  While useful for the analysis of prediction errors due to classification, this scheme is not generally practical because it assumes prior knowledge of the reference energy labels for the test molecules. 
Since the perfect classifier avoids mis-classification of the test data by construction, it should be regarded as a best case scenario for the performance of the clustering/regression/classification approach.

\subsection{The clustering/regression/classification workflow}
Fig.~\ref{fig:diagram} summarizes the combined work flow for training and evaluating a MOB-ML model with clustering. The training involves three steps: First, the training dataset of MOB-ML feature vectors and energy labels are assigned to clusters using the RC method (panel a).  Second, for each cluster of training data, the regression model (LR or GPR) is trained, to enable the prediction of pair correlation energies from the MOB-ML vector.  Third, a classifier is trained from the MOB-ML feature vectors and cluster labels for the training data, to enable the prediction of the cluster assignment from a MOB-ML feature vector.

The resulting MOB-ML model is specified in terms of the method of clustering (RC, for all results presented here), the method of regression (either LR or GPR), and the method of classification (either RFC or the perfect classifier). In referring to a given MOB-ML model, we employ a notation that specifies these options (e.g., RC/LR/RFC or RC/GPR/perfect).

Evaluation of the trained MOB-ML model is explained in Fig.~\ref{fig:diagram}d.  A given molecule is first decomposed into a set of test feature vectors associated with the pairs of occupied MOs.  The classifier is then used to assign each feature vector to an associated cluster.  The cluster-specific regression model is then used to predict the pair correlation energy from the MOB feature vector.  And finally, the pair correlation energies are summed to yield the total correlation energy for the molecule.

To improve the accuracy and reduce the uncertainty in the MOB-ML predictions, ensembles of 10 independent models using the clustering/regression/classification workflow are trained, and the predictive mean and the corresponding standard error of the mean (SEM) are computed by averaging over the 10 models; a comparison between the learning curves\cite{Cortes1994} from a single run and from averaging over the 10 independent models is included in Supporting Information Fig.~S1.
As described here, the predicted correlation energies may exhibit discontinuities as a function of nuclear position, due to changes in the assignment of feature vectors among the clusters; moving forward, this may be avoided with the use of soft (or fuzzy) clustering algorithms.\cite{Baraldi1999}

\section{Results}
\label{sec:results}
\subsection{Clustering and classification in MOB feature space}
We begin by showing that the situation explored in Fig.~\ref{fig:greedy}, in which locally linear clusters overlap, also arises in realistic chemical applications of MOB-ML.  We consider the \qmsbt{} set of drug-like molecules with thermalized geometries, using the diagonal pair correlation energies $\epsd$ computed at the MP2/cc-pVTZ level.
Randomly selecting 1000 molecules for training, we perform RC on the dataset comprised of these energy labels and feature vectors, using  $N=20$ optimized clusters; the sensitivity of RC to the choice of $N$ is examined later.

\begin{figure}[htbp]
\includegraphics[width=0.7\columnwidth]{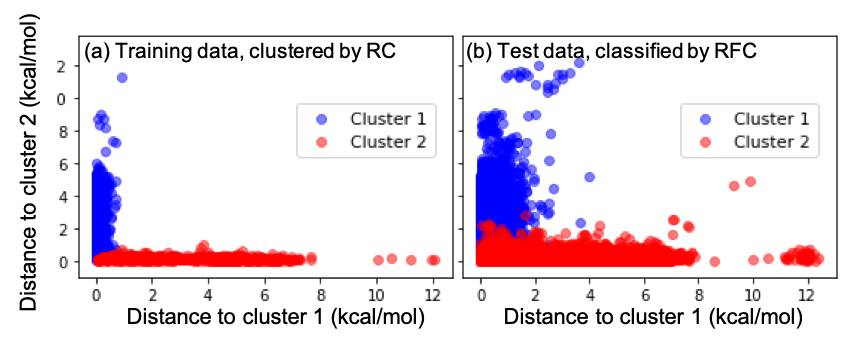}
\caption{
(a) Illustration of the overlap of clusters obtained via RC for the training set molecules from  \qmsbt{}. 
(b) Classification of the datapoints for the remaining test molecules from \qmsbt{}, using RFC.  Distances correspond to the linear regression metric defined in Eq.~\ref{distance_metric}.}
\label{fig:distance}
\end{figure}

In many cases, the resulting clusters are well separated, such that the datapoints for one cluster have small distances (as measured by the linear regression distance metric, Eq.~\ref{distance_metric}) to the cluster which it belongs to and large distances to all other clusters.  However, the clusters can also overlap. Fig.~\ref{fig:distance}a illustrates this overlap for two particular clusters (labeled 1 and 2) obtained from the \qmsbt{} diagonal-pair training data.

Each datapoint assigned to cluster 1 (blue) is plotted according to its distance to both cluster 1 and cluster 2; likewise for the datapoints in cluster 2 (red). The datapoints for which the distances to both clusters approach zero correspond to regions of overlap between the clusters in the high-dimensional space of MOB-ML features, akin to the case shown in Fig.~\ref{fig:greedy}.

Finally, in Fig.~\ref{fig:distance}b, we illustrate the classification of the feature vectors into clusters.
An RFC is trained on the feature vectors and cluster labels for the diagonal pairs of 1000 \qmsbt{} molecules in the training set, and the classifier is then used to predict the cluster assignment for the feature vectors associated with the remaining diagonal pairs of 6211 molecules in \qmsbt{}. 
For clusters 1 and 2, we then analyze the accuracy of the RFC by plotting the linear regression distance for each datapoint to the two clusters, as well as indicating the RFC classification of the feature vector.  Each red datapoint in  Fig.~\ref{fig:distance}b that lies above the diagonal line of reflection is mis-classified into cluster 2, and similarly, each blue datapoint that lies below the line of reflection is mis-classified into cluster 1.  
The figure illustrates that while  RFC is not a perfect means of classification, it is at least qualitatively correct. Later, in the results section, we will analyze the sources of MOB-ML prediction errors due to mis-classification by comparing energy predictions obtained with perfect classification versus RFC.

\begin{figure*}[hbtp]
\includegraphics[width=0.9\textwidth]{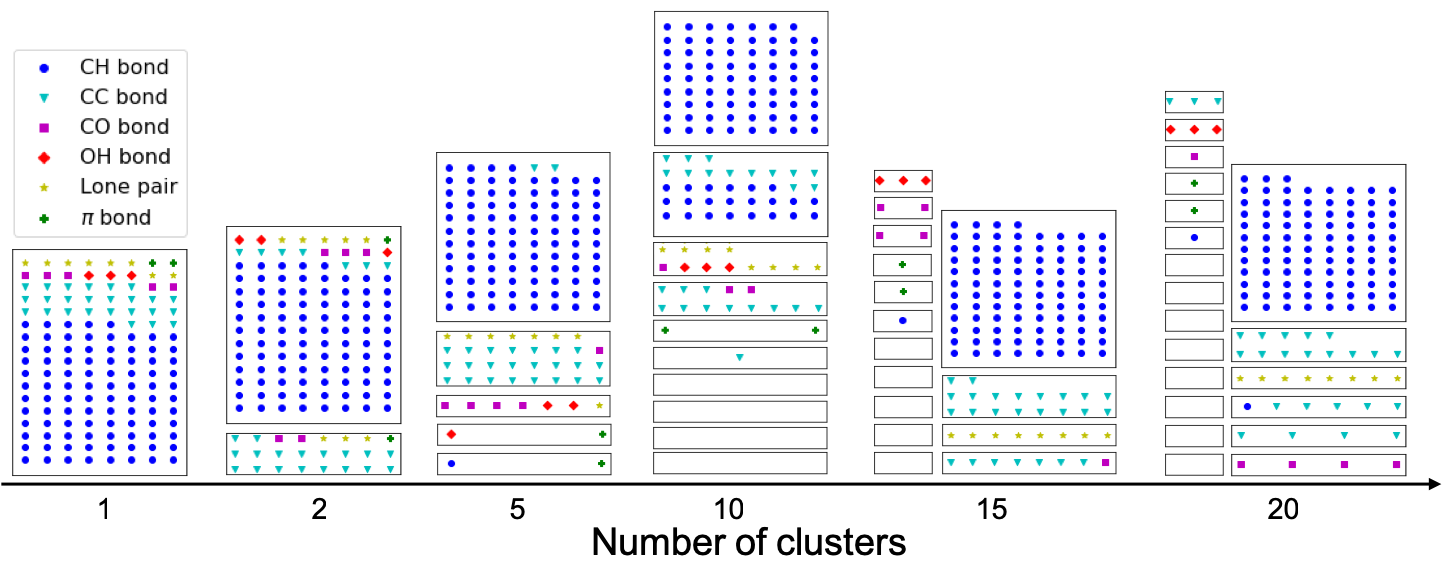}
\caption{
Analyzing the results of clustering/classification in terms of chemical intuition.  
Using a a training set of 500 randomly selected molecules from \qmsbt{}, RC is performed for the diagonal pair correlation energies, $\epsd$, with a range of  cluster numbers, $N$, and for each clustering, an RFC is trained.  Then, the trained classifier is applied to a set of test molecules 
(\ch{CH4}, \ch{C2H6}, \ch{C2H4}, \ch{C3H8}, \ch{CH3CH2OH}, \ch{CH3OCH3}, \ch{CH3CH2CH2CH3}, \ch{CH3CH(CH3)CH3}, \ch{CH3CH2CH2CH2CH2CH2CH3}, \ch{(CH3)3CCH2OH}, and  \ch{CH3CH2CH2CH2CH2CH2OH}) which have chemically intuitive LMO types, as indicated in the legend. The LMOs  are successfully resolved according to type by the classifier as $N$ increases. Empty boxes correspond to clusters into which none of the LMOs from the test set is classified; these are expected since the training set is more diverse than the test set.}
\label{figure:10_mol_diag_cluster}
\end{figure*}

\subsection{Chemically intuitive clusters}
To address this, we employ a training set of 500 randomly selected molecules from \qmsbt{}, and we perform regression clustering for the diagonal pair correlation energies $\epsd$ with a range of total cluster numbers, up to $N=20$.  For each clustering, we then train an RFC.  Finally, each trained RFC is independently applied to a set of test molecules with easily characterized valence molecular orbitals (listed in the caption of Fig.~\ref{figure:10_mol_diag_cluster}), to see how the feature vectors associated with valence occupied LMOs are classified among the optimized clusters.

Figure \ref{figure:10_mol_diag_cluster} presents the results of this exercise, clearly indicating the agreement between chemical intuition and the predictions of the RFC.
As the number of clusters increases, the feature vectors associated with different valence LMO types are resolved into different clusters; and with a sufficiently large number of clusters (15 or 20), each cluster is dominated by a single type of LMO while each LMO type is assigned to a small number of different clusters. The empty boxes in Fig.~\ref{figure:10_mol_diag_cluster} reflect that the training set contains a larger diversity of LMO types than the 11 test molecules, which is expected.
The observed consistency of the clustering/classification method presented here with chemical intuition is of course promising for the accurate local regression of pair correlation energies, which is the focus of the current work; however, the results of Fig.~\ref{figure:10_mol_diag_cluster} also suggest that the clustering/classification of chemical systems in MOB-ML feature space provides a powerful and highly general way of mapping the structure of chemical space for other applications, including explorative or active ML applications.\cite{Browning2017}

\subsection{Sensitivity to the number of clusters} 
We now explore the sensitivity of the MOB-ML clustering/regression/classification implementation to the number of employed clusters.
In particular, we investigate the  mean absolute error (MAE) of the MOB-ML predictions for the diagonal ($\sum_i \epsilon_{ii}$) and off-diagonal ($\sum_{i\neq j} \epsilon_{ij}$) contributions to the total correlation energy, as a function of the number of clusters, $N$, used in the RC.  The MOB-ML models employ linear regression and RFC classification (i.e., the RC/LR/RFC protocol); the training set is comprised of  1000 randomly chosen molecules from \qmsbt{}, and the test set contains the remaining molecules in \qmsbt{}.

\begin{figure}[htbp]
\includegraphics[width=0.6\columnwidth]{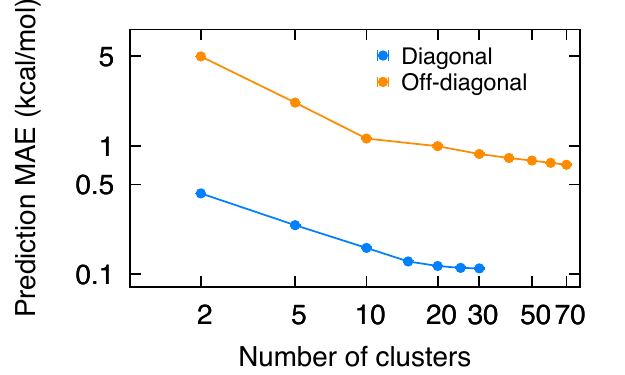}
\caption{
Illustration of the sensitivity of MOB-ML predictions for the diagonal and off-diagonal contributions to the correlation energy for the \qmsbt{} set of molecules, using a subset of 1000 molecules for training and the RC/LR/RFC protocol.
The standard error of the mean (SEM) for the predictions is smaller than the size of the plotted points.}
\label{figure:opt_cluster}
\end{figure}

Figure ~\ref{figure:opt_cluster} presents the result of this calibration study, plotting the  prediction MAE as a function of the number of clusters.
Not surprisingly, the prediction accuracy for both the diagonal and off-diagonal contributions improves with $N$, although it eventually plateaus in both cases.  For the diagonal contributions, the accuracy improves most rapidly up to approximately 20 clusters, in accord with the observations in Fig.~\ref{figure:10_mol_diag_cluster}; and for the off-diagonal contributions, a larger number of clusters is useful for reducing the MAE error, which is sensible given the greater variety of feature vectors that can be created from pairs of LMOs rather than only individual LMOs.  Appealingly, there does not seem to be a strong indication of MAE increases due to ``over-clustering".
While recognizing that the optimal number of clusters will, in general,  depend somewhat on the application and the regression method (i.e., LR versus GPR), the results in Fig.~\ref{figure:opt_cluster} nonetheless provide useful guidance with regard to the appropriate values of $N$. Throughout the remainder of the study, we employ a value of $N=20$ for the MOB-ML prediction of diagonal contributions to the correlation energy and a value of $N=70$ for the off-diagonal contributions; however, we recognize that these choices could be further optimized.

\begin{figure*}[htbp]
\includegraphics[width=0.98\textwidth]{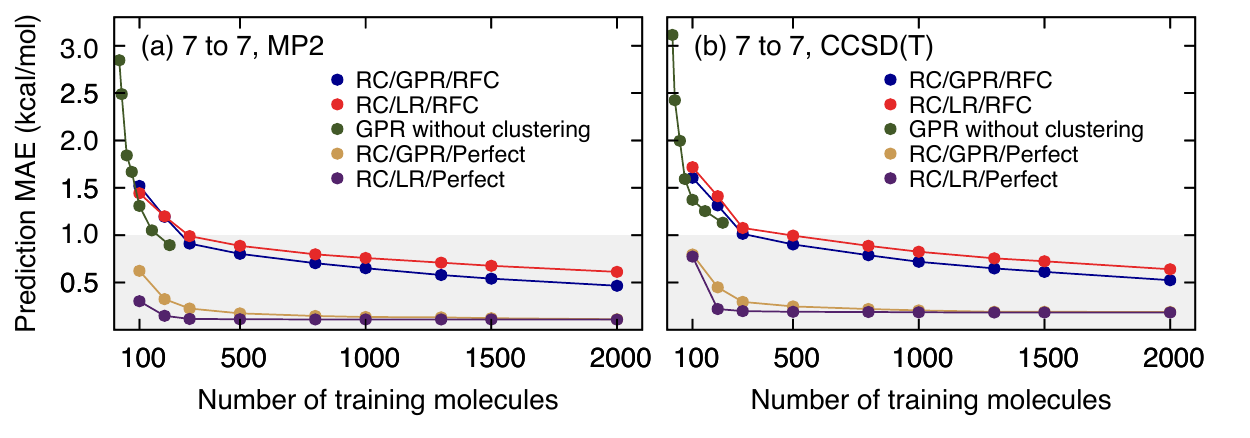}
\caption{
Learning curves for various implementations of MOB-ML applied to (a) MP2/cc-pVTZ and (b) CCSD(T)/cc-pVDZ correlation energies, with the training and test sets corresponding to non-overlapping subsets of the \qmsbt{} set of drug-like molecules with up to heavy seven atoms. 
Results obtained using GPR without clustering (green) are reproduced from Ref.~\citenum{Cheng2019}. 
The gray shaded area corresponds to a MAE of 1 kcal/mol per seven heavy atoms.
The prediction SEM is smaller than the plotted points. The log-log version of this plot is provided in Fig.~S3.}
\label{figure:qm7b_molecule}
\end{figure*}

\begin{figure*}[htbp]
\includegraphics[width=0.58\textwidth]{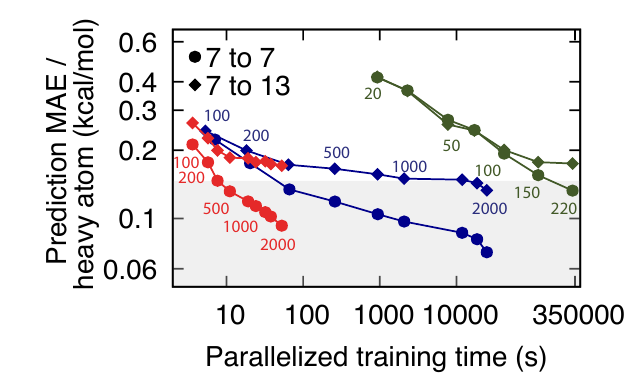}
\caption{
Training costs and transferability of  MOB-ML with clustering (RC/LR/RFC, red; RC/GPR/RFC, blue) and without clustering (green, Ref.~\citenum{Cheng2019}), applied to correlation energies at the MP2/cc-pVTZ level.  Prediction errors are plotted as a function of  wall-clock training time.  
Training sets are comprised of subsets of the \qmsbt{} dataset, with the number of training molecules indicated via datapoint labels. Correlation energy predictions are made for test sets comprised of  the remaining seven-heavy-atom molecules from \qmsbt{} (circles) and the thirteen-heavy-atom molecules from \gdbt{} (diamonds).
Both MAE prediction errors and parallelized wall-clock training times are plotted on a log scale. The gray shaded area corresponds to a MAE of 1 kcal/mol per seven heavy atoms. 
The prediction SEM is smaller than the plotted points. Details of the parallelization and employed computer hardware are described in the text. 
}
\label{figure:7to13}
\end{figure*}

\subsection{Performance and training cost of MOB-ML with RC}
We now investigate the effect of clustering on the accuracy and training costs of MOB-ML for applications to sets of drug-like molecules.
Figure \ref{figure:qm7b_molecule}a presents learning curves (on a linear-linear scale) for various  implementations of MOB-ML applied to MP2/cc-pVTZ correlation energies, with the training and test sets corresponding to non-overlapping subsets of \qmsbt{}. 
In addition to the new results obtained using RC, we include the MOB-ML results from our previous work (GPR without clustering).\cite{Cheng2019}

Figure~\ref{figure:qm7b_molecule}a yields three clear observations. The first is that the use of RC with RFC (i.e., RC/GRP/RFC and RC/LR/RFC) leads to slightly less efficient learning curves than our previous implementation without clustering, at least when efficiency is measured in terms of the number of training molecules.
Both the RC/GPR/RFC and RC/LR/RFC protocols require approximately 300 training molecules to reach the 1 kcal/mol per seven heavy atoms threshold for chemical accuracy employed here, whereas MOB-ML without clustering requires approximately half as many training molecules. The second observation is that the classifier is the dominant source of prediction error in these results. Comparison of results using RFC versus the perfect classifier (which utilizes prior knowledge of the energy labels and this thus not generally practical), reveals a dramatic reduction in the prediction error, regardless of the regression method.  This result indicates that there is potentially much to be gained from the development of improved classifiers for MOB-ML applications.
A third observation is that with a perfect classifier, the LR slightly outperforms GPR, given that the clusters are optimized to be locally linear; however, GPR slightly outperforms LR in combination with the RFC, indicating that GPR is less sensitive to classification error that LR. 

Figure \ref{figure:qm7b_molecule}b presents the corresponding results at the CCSD(T)/cc-pVDZ level of theory. The same trends emerge as the ones at the MP2/cc-pVTZ level of theory. As seen in previous work, the training efficiency of MOB-ML with respect to the size reference dataset is found to be largely insensitive to the level of electronic structure theory.\cite{Welborn2018,Cheng2019} 

Figure \ref{figure:7to13} explores the training costs and transferability of MOB-ML models that employ RC. In all cases, the models are trained on random subsets of molecules from \qmsbt{} with up to seven heavy atoms, and predictions are made either on the remaining molecules of \qmsbt{} (circles) or on the \gdbt{} set (diamonds); it has previously been shown than that MOB-ML substantially outperforms the  FCHL atom-based-feature method in terms of transferability from small to large molecules.\cite{Cheng2019}
The parallelization of the training steps are implemented as follows.
Within the RC step, the LR for each cluster is performed independently on a different core of a 16-core Intel Skylake (2.1 GHz) CPU processor.
Within the regression step, the LR or GPR for each cluster is likewise performed independently on a different core.
For RFC training, we apply parallel 200 cores using the parallel implementation of \textsc{Scikit-learn}, since there are 200 trees. The regression and RFC training are independent of each other and are thus also trivially parallelizable.

Focusing first on the predictions for seven-heavy-atom molecules (circles), it is clear from Fig.~\ref{figure:7to13} that RC leads to large improvements in the efficiency of the MOB-ML wall-clock training costs. Although it requires somewhat more training molecules than MOB-ML without clustering, MOB-ML with clustering enables chemical accuracy to be reached with the training cost reduced by a factor of approximately 4500 for RC/GPR/RFC and of 35000 for RC/LR/RFC. 
Remarkably, for predictions within the \qmsbt{} set, chemical accuracy can be achieved using \lrrfc{} with a wall-clock training time of only 7.7 s. 

Figure~\ref{figure:7to13} also demonstrates the transferability of the MOB-ML models for predictions on the \gdbt{} set of thirteen-heavy-atom molecules (diamonds). In general, it is seen that the degradation in the MAE per atom is  greater for the RC/LR/RFC than for RC/GPR/RFC, due to the previously mentioned sensitivity of LR to classification error.
However, we note that the RC/GPR/RFC enables predictions on \gdbt{} (blue, diamonds) that meet the per-atom threshold of chemical accuracy used here, whereas that threshold was not achievable without clustering (green, diamonds) due to the prohibitive training costs involved.  

The improved efficiency of MOB-ML training with the use of clustering arises from the cubic scaling of standard GPR in terms of training time ($\mathcal{O}(M^3)$, where $M$ is number of training pairs).\cite{rasmussen2006}
Trivial parallelization over the independent regression of the clusters reduces this training time cost to the cube of largest cluster. We note that other kernel-based ML methods with high complexity in training time, like Kernel Ridge Regression,\cite{KRR} would similarly benefit from clustering.
For the \lrrfc{} and \gprrfc{} results presented in Fig.~\ref{figure:7to13}, a breakdown of the training time contributions for each step of the clustering/regression/classification workflow as a function of the size of the training dataset is shown in Fig.~S4; this supporting information figure confirms that the GPR regression dominates the total training (and prediction) costs for the \gprrfc{} implementation, whereas training the RFC dominates the training costs for \lrrfc{}. 
In addition to improved efficiency in terms of training time, clustering also bring benefits in terms of the memory costs for MOB-ML training, due to the quadratic scaling of GPR memory costs in terms of the size of the dataset. 

\begin{figure}[htbp]
\includegraphics[width=0.98\columnwidth]{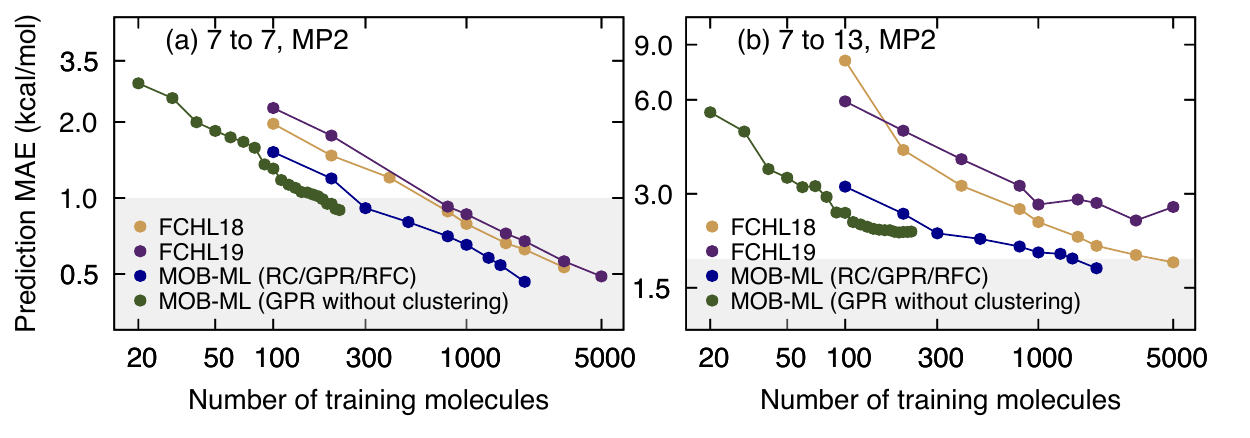}
\caption{Comparison of learning curves for  MP2/cc-pVTZ correlation energies obtained using MOB-ML (with and without clustering) versus  FCHL18 and FCHL19.
Part (a) presents results for which both the training and test sets include molecules from \qmsbt{}, and part (b) presents results for which the training set includes molecules from \qmsbt{} and the test set  includes molecules from \gdbt{}.
The MAE are plotted on a log-log scale as a function of number of training molecules. 
The gray shaded area corresponds to a MAE of 1 kcal/mol per seven heavy atoms.
Results for FCHL18 and FCHL19 were digitally captured from Ref.~\citenum{christensen2019fchl}.
}
\label{figure:fchl_compare}
\end{figure}

Finally, returning to the learning curves, we compare the results for MOB-ML both with and without clustering to recent work\cite{christensen2019fchl} using Faber-Christensen-Huang-Lilienfeld (FCHL) features.  Fig.~\ref{figure:fchl_compare} shows these various learning curves for the MP2/cc-pVTZ correlation energies.  For Fig.~\ref{figure:fchl_compare}a, the training and test sets correspond to non-overlapping subsets of QM7b-T, and  Fig.~\ref{figure:fchl_compare}b  shows the transferability of the same models trained using \qmsbt{} to predict the energies for \gdbt{}.  
Fig.~\ref{figure:fchl_compare}a again shows that MOB-ML RC/GPR/RFC requires slightly more training geometries than MOB-ML without clustering, yet both MOB-ML protocols are more efficient in terms of training data than either the FCHL18\cite{Faber2018} or FCHL19 implementations\cite{christensen2019fchl}. Like MOB-ML with clustering, the FCHL19 implementation was developed to reduce  training times.

\subsection{Capping the cluster size}

Since the parallelized training time for \gprrfc{} is dominated by the GPR regression of the largest cluster (Fig.~S4), a natural question is whether additional computational savings and adequate prediction accuracy could achieved by simply capping the number of datapoints in the largest cluster.
In doing so, we define $S_{\textrm{max}}^{N_{\textrm{cap}}}$ 
to be the number of datapoints in the largest cluster obtained when the RC with the greedy algorithm is applied to a training dataset of $N_{\textrm{cap}}$ molecules from \qmsbt{}.  
Upon specifying $N_{\textrm{cap}}$ (and thus $S_{\textrm{max}}^{N_{\textrm{cap}}}$), the \gprrfc{} implementation is modified as follows.  For a given number of training molecules (which will typically exceed $N_{\textrm{cap}}$), the RC step is performed as normal. However, at the end of the RC step, datapoints for clusters whose size exceeds $S_{\textrm{max}}^{N_{\textrm{cap}}}$ are discarded at random until all clusters contain $S_{\textrm{max}}^{N_{\textrm{cap}}}$ or fewer datapoints.  The GPR and RFC training steps are performed as before, except using this set of clusters that are capped in size.
The precise value of $S_{\textrm{max}}^{N_{\textrm{cap}}}$ will vary slightly depending on which training molecules are randomly selected for training and the convergence of the greedy algorithm, but typical values for $S_{\textrm{max}}^{N_{\textrm{cap}}}$ are $672, 1218, 1863, 3005,$ and $4896$ for $N_{\textrm{cap}}=100, 200, 300, 500$ and $800$, respectively, and those values will be used for the numerical tests presented here. 

\begin{figure}[htbp]
\includegraphics[width=0.98\columnwidth]{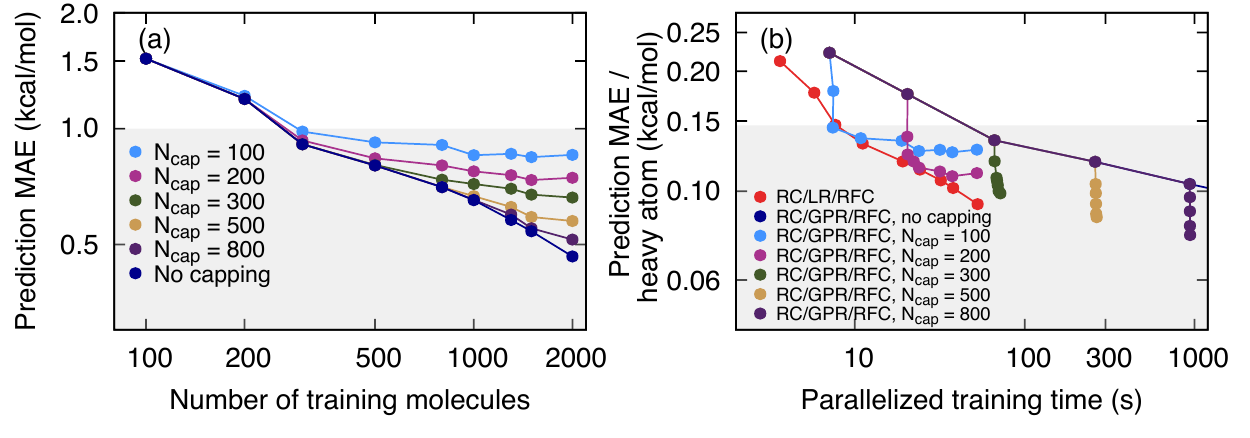}
\caption{
The effect of cluster-size capping on the prediction accuracy and training costs for MOB-ML with RC. Results reported for  correlation energies at the MP2/cc-pVTZ level, with the training and test sets corresponding to non-overlapping subsets of the \qmsbt{} set of drug-like molecules with up to heavy seven atoms.
(a) Prediction MAE versus the number of training molecules, with the clusters capped at various maximum sizes. 
The \gprrfc{} curve without capping is reproduced from Fig.~\ref{figure:qm7b_molecule}a. 
(b) Prediction MAE per heavy atom versus parallelized training time as a function of the number of training molecules, as in Fig.~\ref{figure:7to13}.  
The results for MOB-ML with clustering and without capping cluster size (RC/LR/RFC, red; RC/GPR/RFC, blue) are reproduced from Fig.~\ref{figure:7to13}. 
Also, the results for RC/GPR/RFC  with various capping sizes $N_{\textrm{cap}}$ are shown. 
For part (a), the gray shaded area corresponds to a MAE of 1 kcal/mol, and for part (b), it corresponds to 1 kcal/mol per seven heavy atoms, to provide consistency with preceding figures. 
The prediction SEM is smaller than the plotted points.}
\label{figure:cap}
\end{figure}

Figure~\ref{figure:cap}a demonstrates that capping the maximum cluster size allows for substantial improvements in accuracy when the number of training molecules exceeds $N_{\textrm{cap}}$.  
Specifically, the figure shows the effect of capping on \gprrfc{} learning curves for MP2/cc-pVTZ correlation energies, with the training and test sets corresponding to non-overlapping subsets of \qmsbt{}. 
As a baseline, note that with 100 training molecules, the \gprrfc{} implementation yields a prediction MAE of approximately $1.5$ kcal/mol.  However, if the maximum cluster size is capped at $N_{\textrm{cap}}=100$ and 300 training molecules are employed, then the prediction MAE drops to approximately $1.0$ kcal/mol while the parallelized training cost for  \gprrfc{} will be unchanged so long as it remains dominated by the size of the largest cluster. 
As expected, Fig.~\ref{figure:cap}a shows that the learning curves saturate at higher prediction MAE values when smaller values of $N_{\textrm{cap}}$ are employed.  Nonetheless, the figure demonstrates that if additional training data is available, then the prediction accuracy for MOB-ML with RC can be substantially improved while capping the size of the largest cluster. 

Figure~\ref{figure:cap}b demonstrates the actual effect of capping on the parallelized training time, plotting the prediction MAE versus parallelized training time as a function of the number of training molecules.  For reference, the results obtained using \lrrfc{} and \gprrfc{} without capping are reproduced from Fig.~\ref{figure:7to13}.  
As is necessary, the \gprrfc{} results obtained  with capping exactly overlap those obtained without capping  when the number of training molecules is not greater than $N_{\textrm{cap}}$.  However, for each value of $N_{\textrm{cap}}$, a sharp drop in the prediction MAE is seen when the number of training molecules begins to exceed $N_{\textrm{cap}}$, demonstrating that prediction accuracy can be greatly improved with minimal increase in parallelized training time.  
For example, it is seen that for \gprrfc{} with $N_{\textrm{cap}} =100$, chemical accuracy can be reached with only 7.4 s of parallelized training, slightly less than even \lrrfc{}.  For small values of $N_{\textrm{cap}}$, this prediction MAE eventually levels-off versus the training time, since the RFC training step becomes the dominant contribution to the training time.

\section{Conclusions}

Molecular-orbital-based (MOB) features offer a complete representation for mapping chemical space and a compact representation for evaluating correlation energies.
In the current work, we take advantage of  the intrinsic structure of MOB feature space, which cluster according to types of localized molecular orbitals, as well as the fact that orbital-pair contributions to the correlation energy contributions vary linearly with the MOB features, to overcome a fundamental bottleneck in the efficiency of machine learning (ML) correlation energies. 
Specifically, we introduce a regression clustering (RC) approach in which MOB features and pair correlation energies are clustered according to their local linearity; we then individually regress these clusters and train a classifier for the prediction of cluster assignments on the basis of MOB features.  This combined clustering/regression/classification approach is found to reduce MOB-ML training times by 3-4 orders of magnitude, while enabling prediction accuracies that are substantially improved over that which is possible using MOB-ML without clustering.  The use of a random forest classifier for the cluster assignments, while better than alternatives that were explored,  is found to be the limiting factor in terms of  MOB-ML accuracy within this new approach, motivating future work on improved classifiers.  This work provides a useful step towards that development of accurate, transferable, and scalable quantum ML methods to describe ever-broader swathes of chemical space.

\begin{acknowledgement}
This work emerged from a CMS 273 class project at Caltech that also involved Dmitry Burov, Jialin Song, Ying Shi Teh, and Dr.~Tamara Husch, as well as Professors Kaushik Bhattacharya and Richard Murray; we thank these individuals for their ideas and contributions.
This work is supported by the US Air Force Office of Scientific Research (AFOSR) grant FA9550-17-1-0102.  M.W. acknowledges a postdoctoral fellowship from the Resnick Sustainability Institute.
N.B.K. is supported, in part, by the US National Science Foundation (NSF)
grant DMS 1818977, the US Office of Naval Research (ONR) grant N00014-17-1-2079,
and the US Army Research Office (ARO) grant W911NF-12-2-0022.
Computational resources were provided by the National Energy Research Scientific Computing Center (NERSC), a DOE Office of Science User Facility
supported by the DOE Office of Science under contract DE-AC02-05CH11231.
\end{acknowledgement}

\begin{suppinfo}
Figures in the supplementary information show the effect of averaging over independently trained MOB-ML-models (Fig.~S1), 
the sensitivity of the prediction accuracy to the RC convergence threshold (Fig.~S2), 
learning curves for various implementations of MOB-ML plotted on a log-log scale (Fig.~S3), and a detailed breakdown of the parallelized wall-clock timings (Fig.~S4). Tables in the supporting information provide the numerical data for the plots appearing in the main text.
\end{suppinfo}

\bibliography{main}
\end{document}